\documentclass[conference]{IEEEtran}
\IEEEoverridecommandlockouts
\usepackage{cite}
\usepackage{amsmath,amssymb,amsfonts}
\usepackage{algorithmic}
\usepackage{comment}
\usepackage{tikz}
\usepackage{xcolor}
\usetikzlibrary{positioning,arrows.meta}
\usepackage{graphicx}
\usepackage{textcomp}
\usepackage{xcolor}
\def\BibTeX{{\rm B\kern-.05em{\sc i\kern-.025em b}\kern-.08em
    T\kern-.1667em\lower.7ex\hbox{E}\kern-.125emX}}
\begin{document}


\title{The Internet of Physical AI Agents: Interoperability, Longevity, and the Cost of Getting It Wrong
}

\author{
Roberto Morabito$^\ast$ and Mallik Tatipamula$^\dagger$\\[4pt]
$^\ast$EURECOM, France, $^\dagger$Ericsson Silicon Valley, USA \\
roberto.morabito@eurecom.fr, mallik.tatipamula@ericsson.com
}

\maketitle
\begingroup
\renewcommand\thefootnote{}
\footnotetext{A related version of this work is currently under review for publication in an IEEE magazine.
\\
$^\ast$Roberto Morabito is an Assistant Professor in the Communication Systems Department at EURECOM, France.
\\
$^\dagger$Mallik Tatipamula is CTO at Ericsson, Silicon Valley, with a distinguished 35-year career spanning Nortel, Motorola, Cisco, Juniper, F5 Networks, and Ericsson.
}
\addtocounter{footnote}{-1}

\endgroup

\begin{abstract}
The Internet has evolved by progressively expanding what humanity connects: first computers, then people, and later billions of devices through the Internet of Things (IoT). While IoT succeeded in digitizing perception at scale, it also exposed fundamental limitations, including fragmentation, weak security, limited autonomy, and poor long-term sustainability. Today, advances in edge hardware, sensing, connectivity, and artificial intelligence enable a new phase: the Internet of Physical AI Agents. Unlike IoT devices that primarily sense and report, Physical AI Agents perceive, reason, and act in real time, operating autonomously and cooperatively across safety-critical domains such as disaster response, healthcare, industrial automation, and mobility. However, embedding fast-evolving AI capabilities into long-lived physical infrastructure introduces new architectural risks, particularly around interoperability, lifecycle management, and premature ossification. This article revisits lessons from IoT and Internet evolution, and articulates design principles for building resilient, evolvable, and trustworthy agentic systems. We present an architectural blueprint encompassing agentic identity, secure agent-to-agent communication, semantic interoperability, policy-governed runtimes, and observability-driven governance. We argue that treating evolution, trust, and interoperability as first-class requirements is essential to avoid hard-coding today's assumptions into tomorrow’s intelligent infrastructure, and to prevent the high technical and economic cost of getting it wrong.
\end{abstract}


\section{Introduction}

The Internet has never been just a network. It has always been a story about what humanity decides is worth connecting. It began by linking research computers across a handful of universities. With the Web, it connected people, turning connectivity into a social fabric through email, messaging, and digital communities. The Internet of Things (IoT) extended this reach to billions of devices, embedding connectivity into homes, cities, factories, and vehicles. Each phase expanded the Internet's role, from computation, to communication, to perception.

We now stand at the edge of another transition: the \textbf{\textit{Internet of Physical AI Agents}}.

Unlike IoT devices, which primarily sensed and reported, physical AI agents will \textit{perceive}, \textit{reason}, and \textit{act} in the real world. They are not endpoints streaming telemetry into distant clouds. They are autonomous actors that close the loop between sensing and action. They move, decide, cooperate, and intervene. A drone does not merely observe a wildfire, but it predicts its spread and coordinates suppression. A medical device does not simply report vitals, but it adapts therapy in real time. A robot does not wait for instructions, but it reorganizes production when demand shifts. If IoT connected things, the Internet of Physical AI Agents connects actors.

This transition is not incremental. It represents a shift from a network of sensors to a network of embodied intelligence. From passive data collection to distributed agency. From monitoring the world to shaping it. Yet history offers a cautionary tale.

IoT arrived with enormous promise: smart cities, autonomous infrastructure, self-managing industries. What emerged instead was a fragmented ecosystem of proprietary platforms, battery-hungry devices, insecure endpoints, and complex lifecycle management and integrations. Many deployments stalled. Others delivered data but little action. At planetary scale, even small design mistakes became systemic liabilities.

\begin{figure*}[!t]
    \centering
    \includegraphics[width=1\textwidth]{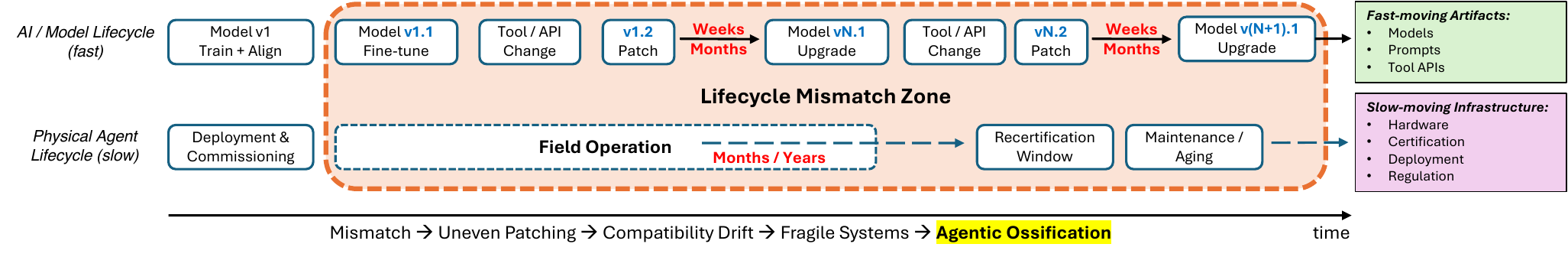}
    \caption{Lifecycle mismatch between fast-moving AI artifacts and slow-moving physical agents increases operational debt and accelerates agentic ossification.}
    \label{fig:lifecycle}
\end{figure*}

We cannot afford to repeat those mistakes. Physical AI agents will operate in safety-critical domains such as healthcare, transportation, disaster response, energy, and manufacturing. Their failures will not merely drop packets. They will endanger lives. Their longevity will be measured not in software release cycles, but in decades of operation in the physical world. And once deployed at scale, their protocols and interfaces will be extremely difficult to evolve.

This is where the next Internet must be built differently.

The Internet succeeded because it was designed for longevity: simple core abstractions, layered architectures, open standards, and continuous evolution without breaking compatibility. At the same time, it also taught us a hard lesson: \textbf{protocols ossify} \cite{honda2011still}. Once deployed at global scale, even flawed mechanisms become nearly impossible to replace. We still carry design decisions from the 1980s in today's networks. Transport Layer Security (TLS) itself, despite multiple iterations, has become a cautionary example of how difficult it is to evolve security once it is deeply embedded in infrastructure \cite{wolsing2019performance}.

The Internet of Physical AI Agents risks an even more dangerous form of ossification: \textbf{agentic ossification}, where identity, communication, safety, autonomy mechanisms, and even AI model interfaces become locked into proprietary silos or prematurely standardized abstractions before their long-term implications are fully understood. Unlike traditional Internet endpoints, Physical AI Agents must operate in safety-critical domains such as healthcare, transportation, disaster response, energy, and manufacturing. Their failures do not merely drop packets; they endanger lives.

This risk is fundamentally driven by a mismatch in lifecycles, illustrated in Figure \ref{fig:lifecycle}. Modern AI artifacts, i.e., models, prompts, reasoning frameworks, and tool APIs, evolve on timescales of days or weeks \cite{tang2020pace}, while physical agents (e.g., robots, medical devices, vehicles, and industrial systems) are deployed, certified, and operated over several months, years or even decades (in a similar fashion as cyber-physical systems \cite{bennaceur2019modelling}). Once intelligence is embedded into such long-lived systems, evolution itself becomes constrained by physical, regulatory, and operational realities. The core systems insight is simple but profound. Intelligence cannot be embedded into long-lived infrastructure without explicitly redesigning the lifecycle through which it evolves. If evolution is not treated as a first-class architectural concern, Physical AI systems will either accumulate unbounded operational risk or become effectively impossible to adapt. These tensions are not only technical but economic. In fact, frequent partial upgrades inflate OPEX through maintenance, recertification, and fleet fragmentation, while accelerated hardware replacement to keep pace with AI evolution drives unsustainable CAPEX.

At the same time, we are already observing a rapid proliferation of agentic protocols, tool APIs, and orchestration frameworks, many driven by corporate platforms rather than by open architectural requirements. This mirrors early Internet dynamics. QUIC, for example, emerged from Google as a highly successful transport protocol, yet required years of standardization and significant redesign before becoming an Internet primitive \cite{langley2017quic}. In the agentic world, the stakes are higher: premature dominance of proprietary protocols could hard-code architectural constraints into the world's intelligent infrastructure before the ecosystem has matured.

If we get this wrong, we will not merely inherit IoT's fragmentation, but we will embed it permanently into the fabric of intelligent systems. And unlike web browsers or smartphones, physical agents cannot simply be replaced every two years.

This article argues that the Internet of Physical AI Agents must therefore be designed from day one for: \emph{(i)} autonomy with reflexes, \emph{(ii)} interoperability without lock-in, \emph{(iii)} security without exception, \emph{(iv)} observability without opacity, and \emph{(v)} evolution without ossification. 

To do so, we must first understand what IoT got right—and where it failed.

\section{Lessons from IoT}

The IoT promised a world in which every object, from thermostats to trucks, would stream data into a global nervous system. Billions of devices now sense and transmit telemetry, digitizing sectors as diverse as logistics, agriculture, energy, and urban planning. By sheer scale alone, IoT changed how the physical world is measured. But it did not change how the world acts.

IoT succeeded in digitizing perception, but it failed to deliver agency. The result is an Internet rich in data but poor in reflexes, i.e., an infrastructure that observes everything and intervenes slowly. In many deployments, value never materialized beyond dashboards and alerts. In others, complexity and maintenance costs quietly erased the promised return on investment. For the Internet of Physical AI Agents, IoT is not a template. It is a warning.

\subsection*{Devices Designed for Demos, Not for Decades}

Early IoT devices were constrained by energy, form factor, and cost. Smart meters required bulky batteries. Environmental sensors needed constant recharging. Industrial deployments demanded frequent maintenance. In theory, devices were meant to be "deploy and forget". In practice, they became "deploy and service".

This created a structural mismatch between ambition and reality. IoT was envisioned as planetary-scale infrastructure, but it was engineered more like consumer electronics. Hardware lifecycles were measured in months. Infrastructure lifecycles were measured in decades.

For physical AI agents, this gap is even more dangerous. A drone fleet, a medical robot, or an autonomous vehicle cannot be treated as disposable hardware. These systems must operate reliably for years in harsh, often inaccessible environments. Their hardware, software, and AI models must evolve gracefully without constant physical intervention. Longevity is not a feature. It is a first-order design requirement.

\subsection*{A Nervous System Without Reflexes}

Most IoT devices were designed as passive sensors. They collected data and transmitted it upstream for processing. Intelligence lived in centralized clouds. Action, when it existed at all, arrived late. In this sense, IoT became the nerve endings of a digital nervous system, but without reflexes.

For time-critical domains, this architecture proved fragile. Fire detection systems could detect smoke but could not respond. Industrial sensors could flag anomalies but could not prevent failures. Smart cities could observe congestion but could not react in real time.

Physical AI agents must close this loop. They must sense, reason, and act locally while coordinating globally. Reflexes must live at the edge. The cloud should provide strategy, learning, and large-scale optimization, but not micromanagement of real-world dynamics. Without reflexes, autonomy is an illusion.

\subsection*{Fragmentation as a Feature, Not a Bug}

IoT was never one Internet. It was thousands of disconnected ecosystems \cite{mumtaz2017massive}. Every vendor shipped its own stack. Every platform defined its own APIs. Devices spoke mutually unintelligible dialects. Smart home systems could not interoperate. Industrial deployments were locked into single vendors. Cross-domain integration was rare and expensive. Fragmentation was not an accident. It was a business model.

The result was a patchwork of silos rather than a unified fabric. Integration costs dominated deployment costs. Innovation slowed as ecosystems hardened around proprietary interfaces.

For Physical AI Agents, fragmentation would be catastrophic. A drone from one vendor must coordinate with a robot from another. A medical agent must trust a hospital system across borders. A disaster response fleet must assemble dynamically from heterogeneous providers. This requires a shared architectural substrate, not a marketplace of incompatible platforms. Interoperability is not optional. It is existential.

\subsection*{Security Treated as a Patch, Not a Foundation}

IoT security was largely retrofitted. Devices shipped with hardcoded credentials, unpatchable firmware, and minimal isolation. At scale, these weaknesses became systemic vulnerabilities. The Mirai botnet demonstrated the consequences \cite{antonakakis2017understanding}. Millions of insecure cameras and DVRs were weaponized into a global attack platform. What looked like isolated design shortcuts became Internet-scale liabilities.

Physical AI agents will be far more powerful than IoT sensors. A compromised thermostat is an inconvenience. A compromised autonomous vehicle is a weapon. Security cannot be a layer added later. It must be embedded in hardware, firmware, models, runtimes, identities, and protocols. Trust must be verifiable. Updates must be cryptographically controlled. Autonomy must be bounded by policy. In an Internet of actors, insecurity is not a bug. It is an existential threat.

\subsection*{Scale Without Value}

IoT was driven by numbers. Tens of billions of devices. Trillions of data points. Zettabytes of telemetry. But scale alone does not create value. Many deployments stalled because they lacked clear return on investment, were too complex to manage, or delivered little beyond monitoring. Data without action proved insufficient. Dashboards without automation became operational burdens.

Physical AI must reverse this logic. Value must come first. Autonomy must deliver measurable outcomes: faster response times, lower costs, safer operations, higher resilience. Scale should follow proven impact, not precede it.

\subsection*{The Hidden Cost: Lifecycle Fragility}

IoT systems, and cyber-physical systems more broadly, carry an implicit long-lived requirement \cite{bennaceur2019modelling}. However, perhaps IoT's most underestimated failure was lifecycle management. Devices were deployed with little consideration for long-term software maintenance, security patching, hardware aging, or evolving protocols. Firmware updates failed. Certificates expired. Backend services changed. Devices were abandoned in the field. At small scale, these failures were tolerable. At global scale, they became unmanageable.

Physical AI agents compound this challenge. Their cognitive core, the AI models themselves, evolves far faster than physical infrastructure. Models are retrained, aligned, compressed, and replaced on timescales measured in weeks. Physical systems evolve on timescales measured in years. Without disciplined lifecycle architecture, agentic systems will drift out of compatibility, security, and policy compliance. What begins as innovation ends as operational debt.

\subsection*{\textbf{Lesson Learned}}

The IoT digitized perception, but not agency. It scaled endpoints, but not trust. It collected data, but delivered limited autonomy. Table \ref{tab:iot_to_physical_ai} summarizes the architectural shift required to move from IoT to the Internet of Physical AI Agents.

\begin{table*}[t]
\centering
\caption{From IoT to the Internet of Physical AI Agents: Architectural Lessons}
\label{tab:iot_to_physical_ai}
\begin{tabular}{p{3cm} p{6cm} p{6cm}}
\hline
\textbf{Dimension} & \textbf{Internet of Things (IoT)} & \textbf{Internet of Physical AI Agents} \\
\hline
Role in the system & Digitized perception & Embodied intelligence \\
Function & Sense and report & Perceive, reason, and act \\
Architecture & Reporting pipeline & Reflexive control system \\
Intelligence & Centralized in the cloud & Distributed across agents and edge \\
Interoperability & Fragmented ecosystems & Open, interoperable fabric \\
Trust model & Retrofitted security & Security by design \\
Lifecycle & Short-lived products & Long-lived infrastructure \\
Evolution & Vendor-driven stacks & Open, evolvable architecture \\
Failure mode & Data without action & Bounded autonomy with accountability \\
Scaling model & Billions of endpoints & Planetary-scale agency \\
Risk & Insecure devices & Unsafe autonomy \\
Long-term threat & Fragmentation & Agentic ossification \\
\hline
\end{tabular}
\end{table*}

The limitations we highlighted should not be read as failures of vision, but as consequences of timing. IoT emerged in an era when low-power AI accelerators did not exist, energy harvesting was immature, connectivity was not deterministic, and edge computing was still in its infancy. Expecting reflexive autonomy from a technology stack built for telemetry is like expecting real-time video conferencing from dial-up modems.

In the same way that Wi-Fi transformed what was once a wired-only Internet, today's advances in AI, sensing, materials, and connectivity finally make it possible to close the loop between perception and action. The Internet of Physical AI Agents is not a correction of IoT, but rather it is its natural successor, enabled by a generation of technologies that simply did not exist before. If IoT taught us how to sense the world, Physical AI must teach us how to responsibly act within it.

\section{From IoT to Physical AI Agents}

Physical AI agents represent the next leap in the Internet's evolution. They are not a new category of connected devices. They are a new class of networked entities:
\begin{itemize}
    \item Where IoT devices sense and transmit, agents perceive, decide, and act.
    \item Where IoT endpoints report, agents intervene.
    \item Where IoT connects things, agents connect intelligence to the physical world.
\end{itemize}

This is not simply an upgrade in capability. It is a shift in the Internet's role from a medium of information exchange to a fabric of distributed cognition and action.

Three properties distinguish Physical AI Agents from the IoT systems that came before.
\\
\textbf{Autonomy}. IoT digitized perception. Agents embody cognition. They observe their environment, reason over goals and constraints, and take action without waiting for centralized control. Autonomy is not an optimization; it is the defining property.
\\
\textbf{Integration.} IoT separated sensing, networking, and intelligence into loosely coupled pipelines. Agents collapse this stack into a closed loop. Sensing, communication, learning, and control operate as a single reflexive system.
\\
\textbf{Collaboration.} IoT devices spoke to clouds. Agents speak to each other. They form distributed multi-agent systems that coordinate, negotiate, and adapt in real time.

In this sense, Physical AI Agents are not endpoints. They are participants.

\subsection*{Digital AI Agents vs. Physical AI Agents}

We are already living in a world populated by digital AI agents: Coding copilots write software, workflow agents orchestrate business processes, trading bots move markets, and customer-service agents negotiate with millions of users every day. These systems live entirely in software, yet they already influence economies, institutions, and knowledge flows. They operate at machine speed, reason over complex state, and interact with both humans and other agents. Physical AI agents extend this intelligence into the real world.

A self-driving vehicle is not just a robot with wheels. It is a mobile decision system operating under uncertainty, interacting with humans, infrastructure, and other autonomous systems. A drone fleet is not a collection of flying cameras, but a distributed sensing-and-action platform. A surgical robot is not a mechanical tool, but a cognitive system embedded in a human feedback loop.

Physical AI agents are then intelligence in motion. They bring all the challenges of digital AI, reasoning, alignment, robustness, security, into domains where mistakes have physical consequences.

\subsection*{The Next Internet Primitive}

Every major phase of the Internet introduced a new primitive: \emph{(i)} the early Internet introduced the host, \emph{(ii)} the Web introduced the resource, \emph{(iii)} mobile introduced the user-in-motion, \emph{(iv)} IoT introduced the sensor.

Physical AI introduces the agent. An agent is not just a device with an IP address. It is an entity with goals, perception, memory, policies, and the ability to act. It has identity, reputation, and accountability. It participates in workflows. It collaborates. It negotiates. It makes decisions. 

Once agents become first-class citizens of the Internet, everything changes:

\begin{itemize}
    \item Routing is no longer just about packets, but it is about \textit{missions}.
    \item Security is no longer just about channels, but it is about \textit{delegated authority}.
    \item Naming is no longer just about endpoints, but it is about \textit{accountable actors}.
    \item Observability is no longer just about traffic, but it is about \textit{decisions}.
\end{itemize}

This is why Physical AI Agents cannot be treated as another vertical industry or another application domain. They require a rethinking of Internet architecture itself.

\subsection*{A Familiar Transition}

This transition mirrors earlier Internet inflection points. From one end we had the Web that moved computing out of research labs and into everyday life. Later, mobile broadband freed the Internet from fixed locations. Finally, cloud computing abstracted infrastructure into programmable services.

Physical AI agents extend intelligence beyond the cloud and into the physical fabric of society. Just as the Internet once connected computers, then people, then things, it is now connecting actors. And just as the Web required new abstractions (URLs, browsers, HTTP), and mobile required new ones (mobility management, radio access, handover), Physical AI will require its own architectural foundations: identity for agents, protocols for coordination, runtimes for autonomy, and governance for safety.

The Internet has gone through this kind of transition before. Each time, success depended not on any single technology, but on architectural discipline: simplicity at the core, intelligence at the edge, layered abstractions, open standards, and neutral governance. The same principles must guide the Internet of Physical AI Agents.

\section{Enablers: Why Now?}

The Internet of Physical AI Agents is not a speculative vision. It is a convergence that is already underway. For decades, the idea of autonomous, cooperative machines operating at global scale remained constrained by fundamental limitations in computing, energy, connectivity, and sensing. IoT emerged before edge intelligence was practical. Robotics matured before reliable, low-latency global connectivity existed. Artificial intelligence advanced rapidly, but for a long time it could only be deployed efficiently inside hyperscale data centers.

Those constraints are now dissolving. A set of independent technology curves, spanning hardware, materials, networking, and AI, have reached a point of alignment. For the first time, it is possible to build physical systems that are compact, intelligent, connected, and autonomous at planetary scale. What once required bespoke infrastructure and centralized control can now be deployed as a distributed, self-organizing fabric of intelligent agents.

\subsection*{Intelligent Devices: From Endpoints to Cognitive Systems}

Modern devices are no longer simple network endpoints. They have evolved into full-fledged computing platforms capable of hosting sophisticated perception and reasoning pipelines.

Smartphones, drones, vehicles, robots, and industrial controllers now integrate GPUs, NPUs, and domain-specific accelerators that enable on-device execution of complex AI models \cite{wang2025empowering}. Computer vision allows agents to interpret their physical surroundings, while speech and language models enable natural interaction with humans and other machines. Planning and control models support decision-making under uncertainty, and reinforcement learning enables continuous adaptation in dynamic environments.

In effect, these systems are no longer merely connected. They are cognitive.

In addition, Edge AI has crossed a critical threshold. Inference workloads that once required cloud-scale infrastructure can now be executed locally within tight latency and energy budgets. This shift enables real-time reflexes, privacy-preserving intelligence, and resilience under intermittent connectivity. It transforms the edge from a passive data source into an active decision-making layer \cite{meuser2024revisiting}.

\subsection*{Generative AI as the Cognitive Layer}

A second inflection point comes from the rise of generative AI. Large language models, multimodal foundation models, and embodied AI systems are redefining how machines reason, plan, and interact with their environment. These models go beyond classification and detection. They interpret context, generate explanations, decompose tasks, and coordinate actions \cite{khoramnejad2025generative}.

For physical agents, generative AI becomes a cognitive layer that enables high-level reasoning and semantic interoperability. It supports natural language interaction with humans, allows agents to reason over goals and constraints, and provides a shared abstraction layer across heterogeneous hardware and software stacks.

Just as the \textit{Web} introduced a universal content layer for the Internet, generative models introduce a universal reasoning layer. When embedded into physical agents, these models transform robots, vehicles, and industrial systems from scripted machines into adaptive collaborators that can understand intent, negotiate trade-offs, and coordinate missions across distributed fleets.

\subsection*{New Materials for Compact, Sustainable Design}

Energy and form factor have long been the Achilles' heel of autonomous systems. IoT deployments were constrained by batteries, maintenance cycles, and physical size, limiting their scalability and operational lifetime.

That constraint is now being relaxed. Advances in meta-materials enable compact, high-performance antennas and sensors, while energy harvesting technologies, spanning solar, thermal, piezoelectric, and RF sources, allow devices to scavenge power from their environment \cite{li2022recent}. Combined with ultra-low-power electronics, these technologies extend operational lifetimes from months to years and, in some cases, enable batteryless operation.

The result is a new class of sustainable agents that can be deployed at scale: lightweight drones with extended endurance, long-lived sensors embedded into infrastructure, wearable and implantable medical devices, and industrial robots capable of operating for years without constant human intervention. These systems are not only intelligent. They are practical.

\subsection*{Next-Generation Connectivity as an Intelligent Fabric}

Connectivity is undergoing a similar transformation. It is no longer defined solely by throughput. Determinism, latency, and reliability have become first-class requirements.

5G URLLC and emerging 6G architectures provide ultra-reliable, low-latency communication with bounded delay, enabling reflex-grade control loops across distributed agents \cite{haque2023survey}. Network slicing allows deterministic performance for safety-critical missions, while distributed cloud and edge computing push computation closer to where action happens.

Integrated Sensing and Communication (ISAC) further collapses radar, perception, and connectivity into a unified stack, turning the network itself into part of the sensing system \cite{liu2022survey}.

In this model, the network is no longer a passive transport substrate. It becomes an active participant in perception, coordination, and control. It carries not only data, but intent, context, and mission state. It enables agents to coordinate in real time with strong performance guarantees.

\subsection*{Closing the Loop}

Individually, each of these advances is significant. Together, they close the loop between perception, reasoning, communication, and action.

Perception is now available at the edge, where physical interaction occurs. Reasoning can be executed in near-real time under strict latency constraints. Communication provides deterministic guarantees for coordination. Action is supported by reflexive control. Learning operates continuously across fleets, enabling collective adaptation.

This is the control loop that IoT could never complete. Where IoT produced telemetry, Physical AI produces decisions. Where IoT required humans in the loop, Physical AI enables autonomy in the loop. Where IoT scaled sensing, Physical AI scales agency.

\subsection*{A System Transition, Not a Product Cycle}

This moment resembles earlier inflection points in the Internet's evolution. The Web did not emerge from a single browser. The mobile Internet was not created by a single phone. Cloud computing was not born from a single data center. Each required a convergence of hardware, software, networking, and economic forces.

The Internet of Physical AI Agents represents the same kind of transition. It is not a product category, a vertical market, or a platform. It is a new layer of civilization's infrastructure, one that embeds intelligence directly into the physical world. For the first time, the technological conditions to build it are finally in place.

\section{Design Principles for the Internet of Physical AI Agents}

The IoT taught us valuable lessons. Rather than treating them merely as cautionary tales, we can reframe them into positive design principles that should guide the Internet of Physical AI Agents toward long-term success. This is not about building better products. It is about building global infrastructure.

Physical AI Agents will operate in safety-critical domains, across borders, across vendors, and across decades. They must survive hardware evolution, model evolution, protocol evolution, and regulatory evolution. They must remain secure long after their original designers have moved on. And they must evolve without ossifying.

If the Internet of Physical AI Agents is to succeed as a planetary-scale system, it must be built on a small set of architectural principles that prioritize longevity, interoperability, safety, and continuous evolution.

\subsection*{Compact, Affordable, and Sustainable by Design}

Where IoT devices were often bulky, battery-limited, and maintenance-heavy, Physical AI Agents must be designed from the ground up for compactness, energy efficiency, and long-term sustainability. Advances in energy harvesting, lightweight metamaterials, and low-power AI accelerators make it possible to deploy agents that operate for extended periods without human intervention. But this must be treated as a design constraint, not a future optimization. A wildfire drone that can patrol autonomously for hours without recharging is not a luxury, but rather a baseline requirement for planetary-scale deployment.

Physical AI Agents will be embedded in forests, oceans, cities, factories, hospitals, and vehicles. Many will operate in environments where replacement is costly or impossible. Longevity must be engineered into their hardware, software, and AI lifecycle from day one. Sustainable design is not only an environmental goal. It is an architectural necessity.

\subsection*{Autonomy with Reflexes, Not Automation Without Agency}

IoT systems collected data but relied on centralized processing and human intervention. Physical AI Agents must embody reflexes. This means embedding perception, reasoning, and control directly into the agent. Computer vision, real-time decision models, and integrated sensing and communication (ISAC) must operate as a closed loop. Agents should act locally while coordinating globally. In precision agriculture, drones should not merely capture crop images and upload them for later analysis. They should autonomously adjust irrigation, apply fertilizer, or trigger pest control in real time. In disaster response, agents should not wait for cloud decisions when every second matters.

Autonomy is not about removing humans. It is about enabling machines to operate safely, responsibly, and predictably when humans cannot be in the loop. Without reflexes, autonomy is an illusion.

\subsection*{Interoperability Through Open Standards}

Fragmentation was one of IoT's most damaging failures. Proprietary APIs, incompatible stacks, and vendor lock-in turned what should have been a global fabric into a marketplace of silos.

The Internet of Physical AI Agents must be built on open, interoperable frameworks from the start. This includes:
\begin{itemize}
    \item \textbf{Agentic identities} that provide universal, verifiable identity across vendors and domains
    \item \textbf{Common semantic formats} for knowledge representation and reasoning
    \item \textbf{Standardized APIs and protocols} for agent-to-agent and agent-to-cloud collaboration
\end{itemize}

At the same time, experience from Internet standardization shows that openness alone is not sufficient. Poorly timed or prematurely scoped standards can be as damaging as proprietary ones, locking in flawed abstractions and inhibiting evolution. Lessons from past IETF standardization efforts highlight how technical, political, and economic pressures can derail otherwise sound designs when consensus is forced too early or driven by narrow interests \cite{welzl2023not}. 

Establishing standards early must therefore be a strong priority, but with an explicit focus on stable abstractions, incremental deployment, and long-term evolvability. Done correctly, this can create an environment in which drones, robots, vehicles, and medical devices—regardless of manufacturer—can cooperate seamlessly in real-world missions without hard-coding today's assumptions into tomorrow's infrastructure.

\subsection*{Security and Trust as First-Class Requirements}

Security must not be retrofitted. It must be intrinsic. Every Physical AI Agent should carry cryptographically verifiable credentials, employ zero-trust architectures, and support secure boot, remote attestation, and authenticated updates. Just as HTTPS became the default for web traffic, secure agent communication must be the default for autonomous systems.

Trustworthy identity is not only a technical requirement. It is a societal prerequisite. A world of autonomous machines requires public confidence. That confidence depends on transparency, accountability, and provable security properties. Without trust, there will be no adoption. Without adoption, there will be no Internet of Physical AI Agents.

\subsection*{High-Value Use Cases Before Hyperscale}

IoT was driven by device counts. Tens of billions of sensors were promised long before value was demonstrated.

Physical AI must invert this logic. Instead of chasing scale for its own sake, it should focus on high-value, mission-critical applications where autonomy delivers measurable benefits: \emph{(i)} Disaster response (wildfires, floods, earthquakes), \emph{(ii)} Healthcare (autonomous intervention devices and monitoring), \emph{(iii)} Industry 5.0 (collaborative robots and adaptive factories), \emph{(iv)} Smart mobility (coordinated fleets of vehicles and drones).

Scaling should follow trust. Trust should follow reliability. And reliability should follow disciplined engineering. Planetary-scale autonomy must be earned.

\subsection*{Reliability, Governance, and Sustainability}

Physical AI Agents must operate at carrier-grade reliability. Failures in healthcare, transportation, or emergency response are unacceptable. Autonomy must degrade gracefully. Systems must be fault-tolerant, self-healing, and resilient to partial failures.

Equally important is governance. The Internet succeeded because its core protocols were not owned by any single company or government. Neutral, transparent institutions such as the IETF and ICANN created global trust and enabled borderless interoperability \cite{ietfAgenticAIStandards}.

The Internet of Physical AI Agents will require similar stewardship: neutral registries, open standards bodies, transparency logs, certification frameworks, and global governance structures that ensure safety, fairness, and accountability.

Finally, sustainability must be designed in from the start. Low-power designs, circular materials, responsible manufacturing, and lifecycle accountability are not optional. Billions of agents will inhabit the physical world. Their environmental footprint must be as carefully engineered as their intelligence.

\section{Toward an Internet of Physical AI Agents: An Architectural Blueprint}

Just as the Internet and the Web succeeded because of a small set of foundational building blocks—such as TCP/IP for networking, HTTP for data exchange, DNS for naming, and the browser as a universal runtime—the Internet of Physical AI Agents requires its own architectural substrate. The longevity of the Internet has been widely attributed to its layered design and to architectural principles that deliberately kept the core simple while allowing innovation and intelligence to evolve at the edges \cite{saltzer1984end}.

This substrate must support autonomy without chaos, intelligence without opacity, scale without fragmentation, and evolution without ossification. It must be simple enough to deploy globally, yet expressive enough to support safety-critical autonomy. Most importantly, it must be designed as long-lived infrastructure, not as a fast-moving software stack.

Rather than defining dozens of protocols and interfaces, we argue for a small number of architectural layers with stable abstractions between them. Each layer should evolve independently, while interoperating through well-defined interfaces, just as the Internet did.

\subsection*{A Layered Architecture for Physical AI Agents}

\begin{figure}[!t]
    \centering
    \includegraphics[width=1\columnwidth]{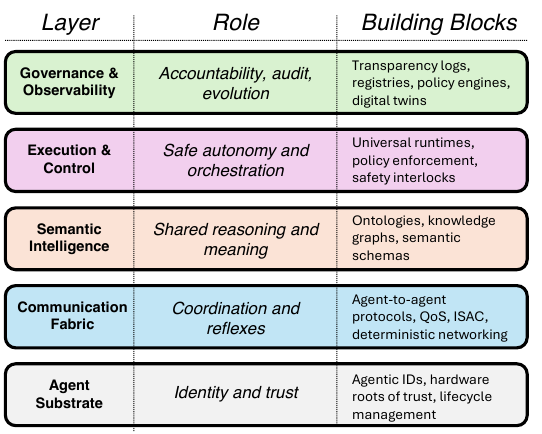}
    \caption{Layered Reference Architecture for the Internet of Physical AI Agents.}
    \label{fig:architecture}
\end{figure}

We envision a layered architecture composed of five foundational layers, as shown in Figure \ref{fig:architecture}.
\begin{figure*}[!t]
    \centering
    \includegraphics[width=1\textwidth]{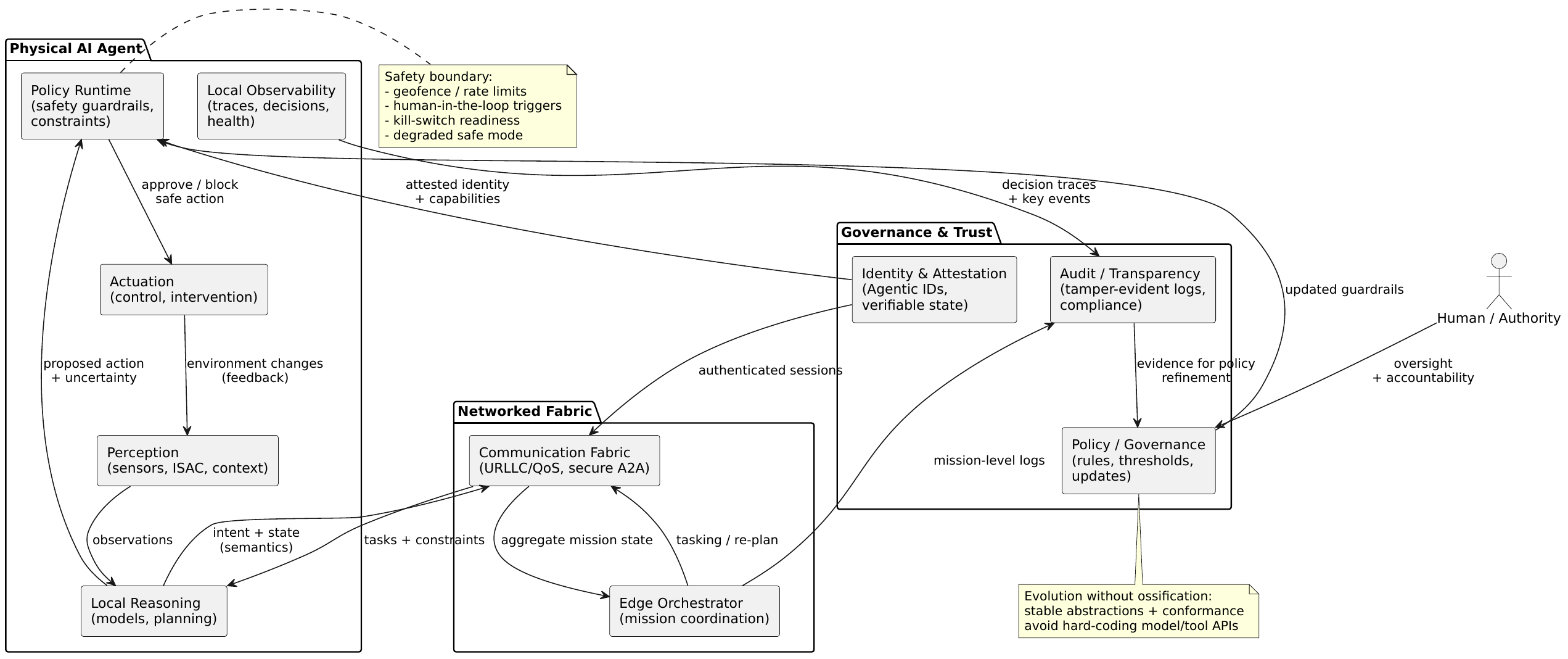}
    \caption{Reference control-loop architecture for Physical AI Agents, integrating local reflexes, fleet-level coordination, and governance feedback under explicit safety and trust constraints.}
    \label{fig:physical_ai}
\end{figure*}
\\
\textbf{Agent Substrate: Identity, Trust, and Lifecycle.} At the foundation lies the agent itself. Every physical AI agent—whether a drone, vehicle, robot, or medical device—must be a first-class Internet entity with a globally verifiable identity. This identity must be cryptographically bound to hardware trust anchors and support the full lifecycle of an agent: registration, delegation, suspension, revocation, and decommissioning. Without secure and persistent identity, there can be no authentication, no authorization, no accountability, and no trust. This layer establishes the notion of the agent as an accountable actor rather than a disposable endpoint.
\\
\textbf{Communication Fabric: Coordination and Reflexes.} Above identity sits the communication fabric. Agents must communicate directly with each other and with infrastructure, not only with centralized clouds. This fabric must support mutual authentication, encrypted multimodal exchange (video, LiDAR, telemetry), quality-of-service guarantees, and degraded-mode operation under failure. This is the equivalent of TCP/IP and HTTP for physical AI systems: a lightweight, low-latency, safety-aware coordination layer that enables distributed reflexes. In this model, the network is not merely a transport substrate. It is an integral part of the control loop.
\\
\textbf{Semantic Intelligence: Meaning, Not Just Data.} IoT systems exchanged data. Physical AI agents must exchange meaning. Agents must reason over shared representations of the world, using common ontologies, knowledge graphs, and semantic schemas. A drone should communicate “probable wildfire, 87\% confidence, wind-driven spread toward sector B” rather than simply reporting a temperature anomaly. This layer provides the shared cognitive substrate that allows heterogeneous fleets to reason coherently across domains, vendors, and geographies. Without semantic interoperability, collaboration collapses into translation overhead.
\\
\textbf{Execution \& Control: Runtimes, Policies, and Safety.} Physical AI needs a universal execution environment, just as the Web needed the browser. This layer provides secure, sandboxed runtimes for deploying agent logic, enforcing policies, and orchestrating fleets. It governs energy budgets, geofencing, safety constraints, model execution, and actuation boundaries. It supports confidential computing, remote attestation, and verifiable execution. It enforces fail-safe interlocks, human-in-the-loop escalation, kill-switches, and degraded safe modes. This is where autonomy becomes governable.
\\
\textbf{Governance \& Observability: Accountability and Evolution.} Finally, autonomy must be observable, auditable, and governable. This layer provides:
\begin{itemize}
    \item Telemetry schemas for agent health and decisions
    \item Distributed tracing across multi-agent missions
    \item Tamper-evident logs for audit and forensics
    \item Digital twins for simulation and patch validation
    \item Service-level objectives for autonomy and safety
    \item Policy feedback loops with human oversight
\end{itemize}

Just as importantly, this layer provides neutral stewardship: registries, transparency logs, conformance testing, and multi-stakeholder governance bodies similar to the IETF and ICANN. This is what prevents fragmentation, lock-in, and agentic ossification.

\subsection*{Why This Matters}

This architecture is intentionally minimal. It avoids monolithic stacks, vendor-specific frameworks, and hard-coding model interfaces that will inevitably evolve. Instead, it provides stable abstractions around identity, communication, semantics, execution, and governance—allowing models, tools, and hardware to evolve independently without breaking the system. This is how the Internet survived five decades of technological disruption. It is also how the Internet of Physical AI Agents can avoid premature ossification.

\subsection*{From Architecture to Operation.}

Figure \ref{fig:physical_ai} translates the architectural blueprint into an operational perspective, showing how a Physical AI Agent behaves as a closed-loop system embedded within a broader networked and governance fabric. At the agent level, autonomy is realized through a tight local loop in which perception informs local reasoning, proposed actions are validated against explicit policy constraints, and actuation closes the loop through interaction with the physical environment. This reflexive behavior is complemented by a coordination loop, where agents exchange intent and state over a deterministic communication fabric and rely on edge orchestration to support task allocation, collective adaptation, and mission-level re-planning. Importantly, identity, attestation, observability, and audit are not external services layered on top of autonomy, but integral elements of the control loop itself. Decision traces and key events feed governance mechanisms that refine policies, thresholds, and constraints over time, with human oversight remaining explicit rather than implicit. By separating fast local reflexes, distributed coordination, and slower governance feedback, this control-loop abstraction enables Physical AI Agents to operate autonomously without losing accountability, and to evolve without freezing interfaces prematurely. The following case studies illustrate how this operational model materializes across diverse, safety-critical domains.

\section{Case Studies: Physical AI Agents in Action}

The Internet of Physical AI Agents is not a distant vision. Early forms of it are already emerging across disaster response, healthcare, industry, and mobility. These domains expose the limitations of IoT-style architectures and demonstrate why reflexive, collaborative, and accountable autonomy is becoming a necessity rather than a luxury.

Each case study illustrates a different facet of the agentic Internet: perception and action at the edge, real-time coordination, semantic interoperability, and system-level resilience.

\subsection*{Wildfire Response: Reflexes at Planetary Scale}

Wildfires are among the most time-critical disasters humanity faces. A fire can double in size in minutes. Smoke spreads faster than human response teams can mobilize. Terrain, wind, and vegetation interact in unpredictable ways.

In an IoT-centric model, forests are equipped with sensors that detect smoke, temperature, and humidity. These sensors report telemetry to centralized servers, where analytics systems attempt to infer fire conditions and alert emergency services. By the time a response is triggered, valuable time has already been lost.

In a Physical AI Agent world, wildfire response becomes a distributed reflex system. Autonomous drones patrol forest corridors, continuously mapping terrain using thermal vision, LiDAR, and atmospheric sensing. When smoke is detected, nearby agents collaborate to triangulate the ignition point, predict fire spread using local wind and vegetation models, and coordinate suppression strategies. Ground robots deploy fire retardants. Aerial agents perform targeted water drops. Edge compute nodes run predictive simulations, while cloud models provide strategic forecasts. Each agent operates autonomously, yet cooperates through a shared semantic model of the environment. Identity, trust, and policy ensure that only authorized responders participate. Observability pipelines provide real-time audit trails for public agencies.

\begin{figure}[!t]
    \centering
    \includegraphics[width=1\columnwidth]{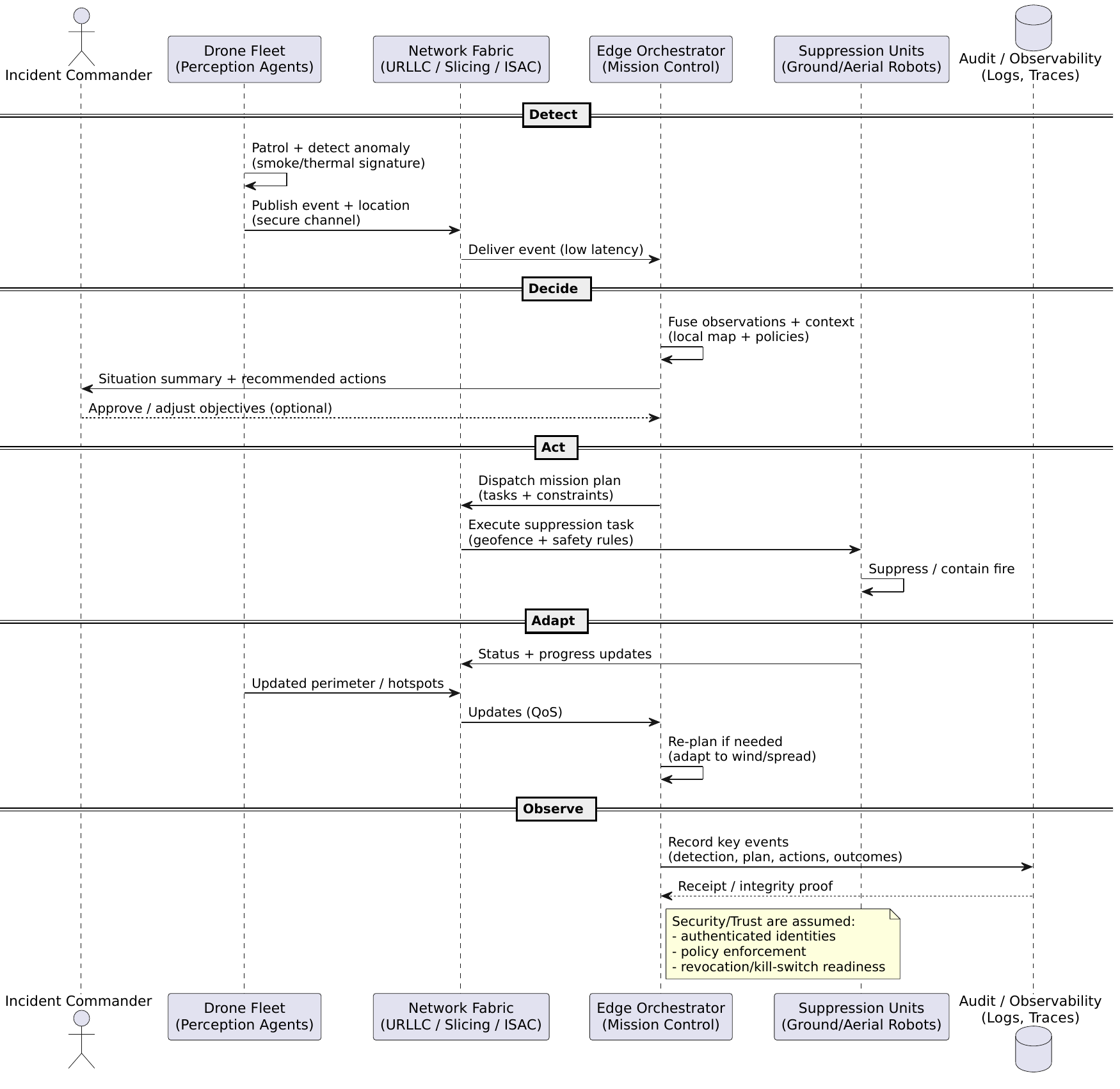}
    \caption{Distributed wildfire response system based on Physical AI Agents. Autonomous drones detect and classify fire events, coordinate through an edge mission orchestrator, and trigger suppression actions over a deterministic communication fabric with identity, policy, and audit controls.}
    \label{fig:w}
\end{figure}

Figure \ref{fig:w} shows how a wildfire response mission is executed as a distributed reflex system, where perception, reasoning, coordination, and intervention operate as a closed loop across autonomous drones, edge orchestrators, and suppression robots.

\begin{figure}[!t]
    \centering
    \includegraphics[width=1\columnwidth]{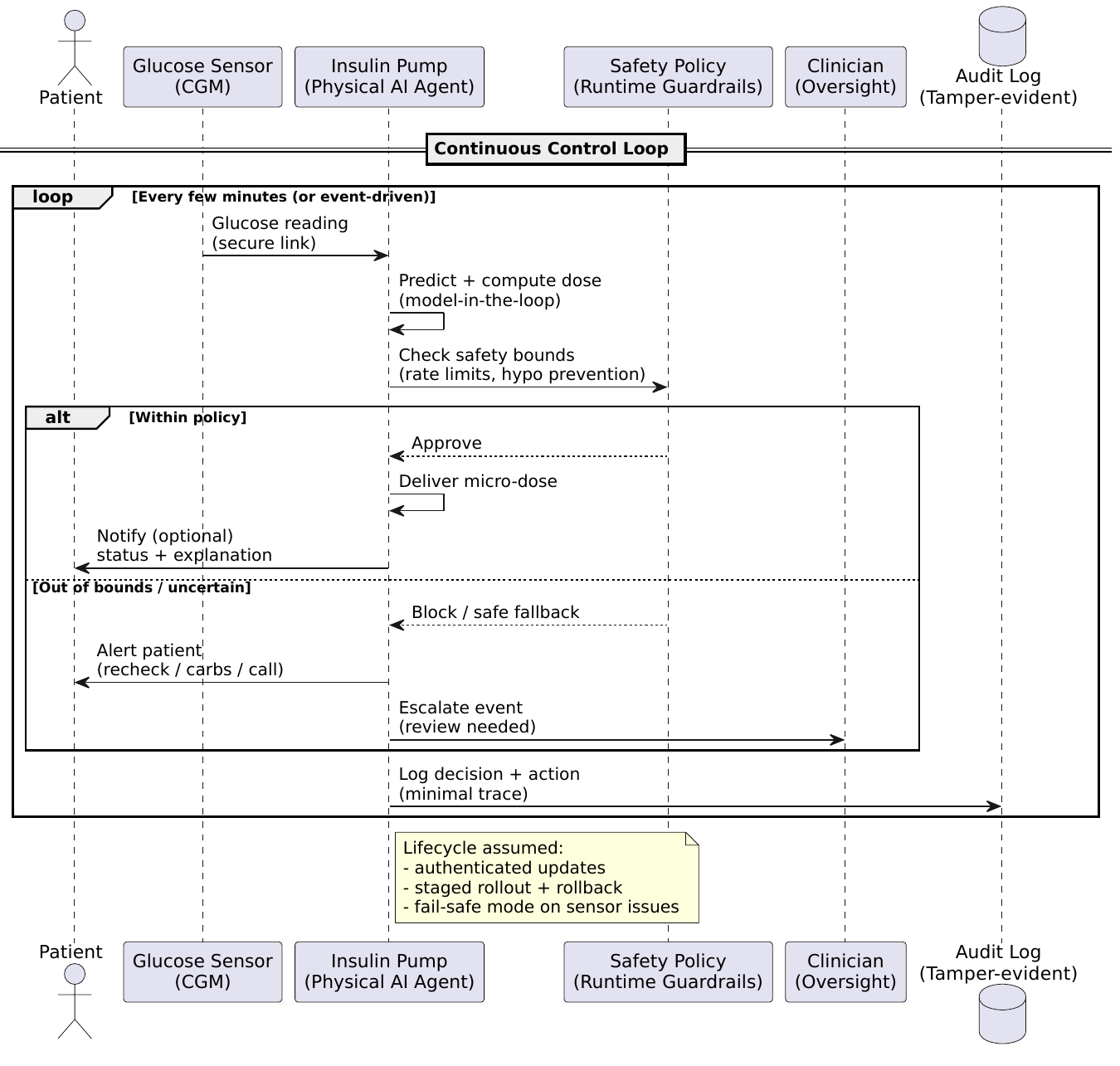}
    \caption{Closed-loop insulin delivery system implemented as a Physical AI Agent, integrating continuous sensing, on-device prediction and control, safety policy enforcement, secure lifecycle management, and auditability.}
    \label{fig:h}
\end{figure}

\subsection*{Healthcare: Closed-Loop Medicine}

Healthcare is fundamentally a control problem. It is about sensing, diagnosis, intervention, and continuous adjustment. Yet most medical IoT systems stop at monitoring.

Wearables stream heart rate, glucose levels, or oxygen saturation to dashboards. Clinicians interpret trends and decide when to intervene. In emergency scenarios, minutes matter. In chronic care, delays accumulate into systemic risk.

Physical AI Agents enable closed-loop medicine. Consider an autonomous insulin delivery system. A wearable sensor continuously monitors glucose. An embedded AI model predicts metabolic response. The pump adjusts dosage in real time. The agent reasons over patient-specific profiles, historical data, and safety constraints. When uncertainty increases, it escalates to clinicians. When anomalies arise, it triggers intervention.

In surgical robotics, agentic systems adapt to patient vitals during procedures. In intensive care, agents manage ventilators, infusion pumps, and monitoring systems as a coordinated ensemble. These systems are not tools. They are clinical collaborators. Identity and authentication ensure that only certified devices and clinicians participate. Semantic models encode medical knowledge. Runtimes enforce safety policies. Observability provides auditability for regulators and hospitals.

Figure \ref{fig:h} illustrates how a closed-loop insulin delivery system can be implemented as a safety-governed Physical AI Agent, where prediction, control, policy enforcement, and auditability are integrated into a single accountable medical device.

\subsection*{Industry 5.0: Adaptive, Self-Organizing Factories}

Industrial IoT transformed factories into sensor-rich environments. Machines report status. Predictive maintenance reduces downtime. Dashboards visualize production flows. But decision-making remains largely centralized and manual.

Physical AI Agents enable factories that reason and adapt as distributed systems. Collaborative robots negotiate task allocation in real time. When a machine fails, nearby agents reconfigure workflows. When demand shifts, production lines reorganize themselves. Energy consumption is optimized dynamically. Safety systems respond instantly to hazardous conditions.

A factory becomes a multi-agent system. Each robot has identity, policies, and goals. They coordinate through agent-to-agent protocols. Semantic knowledge describes production processes and safety constraints. Runtimes enforce operational boundaries. Digital twins simulate changes before deployment. This is not automation. It is industrial cognition.

\subsection*{Urban Mobility: Coordinated Intelligence on the Move}

Urban mobility is one of the most complex systems humanity operates. Millions of vehicles, pedestrians, traffic signals, and infrastructure components interact in dense, dynamic environments. IoT enabled ride-hailing, fleet tracking, and traffic analytics. But coordination remains limited.

Physical AI Agents enable cities that move as coherent systems. Autonomous vehicles negotiate routes collaboratively to minimize congestion and energy consumption. Drone fleets manage logistics and emergency delivery. Traffic infrastructure participates as an intelligent agent, coordinating signals with vehicle flows. Public transport adapts in real time to demand. Each agent reasons locally while sharing intent and context globally. Identity and trust ensure safe interaction. Policies enforce city regulations. Observability enables accountability. Mobility becomes a distributed control system rather than a collection of independent actors.

\subsection*{What These Systems Have in Common}

Across these domains, a common architecture emerges. Agents possess verifiable identities. They \emph{(i)} communicate through deterministic and secure fabrics, \emph{(ii)} share semantic models of their environment, 
\emph{(iii)} execute under policy-governed runtimes, and \emph{(iv)} operate under continuous observation and audit.

This is not a collection of applications. It is a new Internet layer. The Internet of Physical AI Agents is already taking shape. The question is not whether it will exist, but whether it will be built as an open, trustworthy, and evolvable infrastructure—or as a fragmented patchwork of proprietary platforms. The answer will determine whether autonomy becomes a public good or a private asset.

\section{Conclusion}

The IoT connected the world's sensors. The Internet of Physical AI Agents will connect the world’s intelligence. By learning from IoT’s limitations—bulky devices, lack of reflexes, fragmentation, weak security—we can design a new system that is compact, responsive, interoperable, and sustainable. With advances in device hardware, ISAC, smart materials, and intelligent connectivity, the tools to realize this vision are already here. The stakes are high. Physical AI Agents will operate in healthcare, transportation, industry, and disaster response—domains where failures cost lives. If built on open, secure, and interoperable foundations, they can unlock unprecedented benefits. If trapped in silos and walled gardens, they risk amplifying IoT's failures at planetary scale. The Internet's history gives us the blueprint. This time, we must build it right.


\bibliographystyle{IEEEtran}
\bibliography{references}

@inproceedings{langley2017quic,
  title={The quic transport protocol: Design and internet-scale deployment},
  author={Langley, Adam and Riddoch, Alistair and Wilk, Alyssa and Vicente, Antonio and Krasic, Charles and Zhang, Dan and Yang, Fan and Kouranov, Fedor and Swett, Ian and Iyengar, Janardhan and others},
  booktitle={Proceedings of the conference of the ACM special interest group on data communication},
  pages={183--196},
  year={2017}
}

@inproceedings{honda2011still,
  title={Is it still possible to extend TCP?},
  author={Honda, Michio and Nishida, Yoshifumi and Raiciu, Costin and Greenhalgh, Adam and Handley, Mark and Tokuda, Hideyuki},
  booktitle={Proceedings of the 2011 ACM SIGCOMM conference on Internet measurement conference},
  pages={181--194},
  year={2011}
}

@inproceedings{wolsing2019performance,
  title={A performance perspective on web optimized protocol stacks: TCP+ TLS+ HTTP/2 vs. QUIC},
  author={Wolsing, Konrad and R{\"u}th, Jan and Wehrle, Klaus and Hohlfeld, Oliver},
  booktitle={Proceedings of the 2019 Applied Networking Research Workshop},
  pages={1--7},
  year={2019}
}

@article{tang2020pace,
  title={The pace of artificial intelligence innovations: Speed, talent, and trial-and-error},
  author={Tang, Xuli and Li, Xin and Ding, Ying and Song, Min and Bu, Yi},
  journal={Journal of Informetrics},
  volume={14},
  number={4},
  pages={101094},
  year={2020},
  publisher={Elsevier}
}

@article{mumtaz2017massive,
  title={Massive Internet of Things for industrial applications: Addressing wireless IIoT connectivity challenges and ecosystem fragmentation},
  author={Mumtaz, Shahid and Alsohaily, Ahmed and Pang, Zhibo and Rayes, Ammar and Tsang, Kim Fung and Rodriguez, Jonathan},
  journal={IEEE industrial electronics magazine},
  volume={11},
  number={1},
  pages={28--33},
  year={2017},
  publisher={IEEE}
}

@inproceedings{bennaceur2019modelling,
  title={Modelling and analysing resilient cyber-physical systems},
  author={Bennaceur, Amel and Ghezzi, Carlo and Tei, Kenji and Kehrer, Timo and Weyns, Danny and Calinescu, Radu and Dustdar, Schahram and Hu, Zhenjiang and Honiden, Shinichi and Ishikawa, Fuyuki and others},
  booktitle={2019 IEEE/ACM 14th International Symposium on Software Engineering for Adaptive and Self-Managing Systems (SEAMS)},
  pages={70--76},
  year={2019},
  organization={IEEE}
}

@inproceedings{antonakakis2017understanding,
  title={Understanding the mirai botnet},
  author={Antonakakis, Manos and April, Tim and Bailey, Michael and Bernhard, Matt and Bursztein, Elie and Cochran, Jaime and Durumeric, Zakir and Halderman, J Alex and Invernizzi, Luca and Kallitsis, Michalis and others},
  booktitle={26th USENIX security symposium (USENIX Security 17)},
  pages={1093--1110},
  year={2017}
}

@article{wang2025empowering,
  title={Empowering edge intelligence: A comprehensive survey on on-device ai models},
  author={Wang, Xubin and Tang, Zhiqing and Guo, Jianxiong and Meng, Tianhui and Wang, Chenhao and Wang, Tian and Jia, Weijia},
  journal={ACM Computing Surveys},
  volume={57},
  number={9},
  pages={1--39},
  year={2025},
  publisher={ACM New York, NY}
}

@article{meuser2024revisiting,
  title={Revisiting edge ai: Opportunities and challenges},
  author={Meuser, Tobias and Lov{\'e}n, Lauri and Bhuyan, Monowar and Patil, Shishir G and Dustdar, Schahram and Aral, Atakan and Bayhan, Suzan and Becker, Christian and De Lara, Eyal and Ding, Aaron Yi and others},
  journal={IEEE Internet Computing},
  volume={28},
  number={4},
  pages={49--59},
  year={2024},
  publisher={IEEE}
}

@article{khoramnejad2025generative,
  title={Generative AI for the optimization of next-generation wireless networks: Basics, state-of-the-art, and open challenges},
  author={Khoramnejad, Fahime and Hossain, Ekram},
  journal={IEEE Communications Surveys \& Tutorials},
  year={2025},
  publisher={IEEE}
}

@article{li2022recent,
  title={Recent advances in new materials for 6G communications},
  author={Li, Zechen and Pan, Jialiang and Hu, Haowen and Zhu, Hongwei},
  journal={Advanced Electronic Materials},
  volume={8},
  number={3},
  pages={2100978},
  year={2022},
  publisher={Wiley Online Library}
}

@article{liu2022survey,
  title={A survey on fundamental limits of integrated sensing and communication},
  author={Liu, An and Huang, Zhe and Li, Min and Wan, Yubo and Li, Wenrui and Han, Tony Xiao and Liu, Chenchen and Du, Rui and Tan, Danny Kai Pin and Lu, Jianmin and others},
  journal={IEEE Communications Surveys \& Tutorials},
  volume={24},
  number={2},
  pages={994--1034},
  year={2022},
  publisher={IEEE}
}

@article{haque2023survey,
  title={A survey of scheduling in 5G URLLC and outlook for emerging 6G systems},
  author={Haque, Md Emdadul and Tariq, Faisal and Khandaker, Muhammad RA and Wong, Kai-Kit and Zhang, Yangyang},
  journal={IEEE access},
  volume={11},
  pages={34372--34396},
  year={2023},
  publisher={IEEE}
}

@inproceedings{welzl2023not,
  title={How not to IETF: Lessons learned from failed standardization attempts},
  author={Welzl, Michael and Ott, J{\"o}rg and Perkins, Colin and Islam, Safiqul and Kutscher, Dirk},
  booktitle={2023 IEEE International Conference on Pervasive Computing and Communications Workshops and other Affiliated Events (PerCom Workshops)},
  pages={427--432},
  year={2023},
  organization={IEEE}
}

@article{saltzer1984end,
  title={End-to-end arguments in system design},
  author={Saltzer, Jerome H and Reed, David P and Clark, David D},
  journal={ACM Transactions on Computer Systems (TOCS)},
  volume={2},
  number={4},
  pages={277--288},
  year={1984},
  publisher={Acm New York, NY, USA}
}

@online{ietfAgenticAIStandards,
  author       = {Cullen Jennings},
  title        = {Agentic AI communications: Identifying the standards we need},
  year         = {2026},
  month        = {January},
  url          = {https://www.ietf.org/blog/agentic-ai-standards/},
  organization = {Internet Engineering Task Force (IETF)},
  note         = {Accessed: 2026-01-28}
}

\end{document}